\documentclass[a4paper,11pt]{article}

\usepackage{jcappub} 

\usepackage[T1]{fontenc}

\usepackage{graphicx} 
\usepackage{amsmath}
\usepackage{amssymb}

\usepackage{slashed}
\usepackage{subcaption}
\usepackage{xcolor}

\def\be{\begin{equation}}
\def\ee{\end{equation}}
\def\ba{\begin{eqnarray}}
\def\ea{\end{eqnarray}}

\def\bdm{\begin{displaymath}}
\def\edm{\end{displaymath}}

\newcommand{\bea}{\begin{eqnarray}}
\newcommand{\eea}{\end{eqnarray}}

\newcommand{\bi}{\begin{itemize}}
\newcommand{\ei}{\end{itemize}}

\newcommand{\beq}{\begin{equation}}
\newcommand{\eeq}{\end{equation}}
\newcommand{\beqa}{\begin{eqnarray}}
\newcommand{\eeqa}{\end{eqnarray}}

\def\ltap{\ \raise.3ex\hbox{$<$\kern-.75em\lower1ex\hbox{$\sim$}}\ }
\def\gtap{\ \raise.3ex\hbox{$>$\kern-.75em\lower1ex\hbox{$\sim$}}\ }
\def\gl{\ \raise.5ex\hbox{$>$}\kern-.8em\lower.5ex\hbox{$<$}\ }
\def\roughly#1{\raise.3ex\hbox{$#1$\kern-.75em\lower1ex\hbox{$\sim$}}}

\begin{document}

\title{Radiation Exposure from the Dark}

\author[a]{Florian Niedermann}
\author[b]{Martin S. Sloth}
\affiliation[a]{Nordita, KTH Royal Institute of Technology and Stockholm University, Hannes Alfv\'ens v\"ag 12, SE-106 91 Stockholm, Sweden}
\affiliation[b]{Universe-Origins, University of Southern Denmark, Campusvej 55, 5230 Odense M, Denmark}%
\emailAdd{florian.niedermann@su.se}
\emailAdd{sloth@sdu.dk}

\abstract{
We explore the possibility that exotic forms of dark matter could expose humans on Earth or on prolonged space travel to a significant radiation dose. The radiation exposure from dark matter interacting with nuclei in the human body is generally assumed to be negligible compared to other sources of background radiation. However, as we discuss here, current data allow for dark matter models where this is not necessarily true. In particular, if dark matter is heavier and more strongly interacting than weakly interacting massive particle dark matter, it could act as ionizing radiation and deposit a significant amount of radiation energy in all or part of the human population, similar to or even exceeding the known radiation exposure from other background sources. Conversely, the non-observation of such an exposure can be used to constrain this type of  heavier and more strongly interacting dark matter. We first consider the case where dark matter scatters elastically and identify the relevant parameter space in a model-independent way. We also discuss how previous bounds from cosmological probes, as well as atmospheric and space-based detectors, might be avoided, and how a re-analysis of existing radiation data, along with a simple experiment monitoring ionizing radiation in space with a lower detection threshold, could help constrain part of this parameter space. We finally propose a hypothetical dark matter candidate that scatters inelastically and argue that, in principle, one per mille of the Earth's population could attain a significant radiation dose from such a dark matter exposure in their lifetime.}

\date{\today}

\maketitle
\section{Introduction}

The use of biological dark matter detectors, including the use of humans as dark matter detectors, is a new research area still in the pioneering stage, initiated by a group of researchers who proposed to use single-stranded DNA and enzymatic reactions for dark matter detection \cite{Drukier:2012hj,Drukier:2014rea}. More recently, Sidhu, Scherrer, and Starkman studied the possibility that collisions of dark matter with the human body could result in serious injury or death \cite{Sidhu:2019oii}. They found that this is possible if dark matter is much heavier and more strongly interacting than usually believed \cite{Starkman:1990nj}. The idea that dark matter is heavy and strongly interacting has gained increasing momentum in recent years as lighter and weaker interacting dark matter candidates, such as Weakly Interacting Massive Particle (WIMP) dark matter, are getting increasingly constrained \cite{Akerib:2017kat,Aprile:2018dbl}. The recent pioneering work by Sidhu, Scherrer, and Starkman has, in their own words, opened a new window on dark matter: the human body as a dark matter detector.

In this paper, we will study the possibility that if dark matter is heavier and interacts stronger than the WIMP type dark matter but is not heavy and strongly interacting enough to cause serious injury or instant death, it could still have significant effects on humans in terms of increased radiation exposure. 

Dark Matter is usually assumed not to be dangerous to the human body since WIMP dark matter, the favored dark matter candidate among physicists, will scatter elastically with nucleons with a very small cross-section. The recoil energy in the elastic scattering process scales as $E_R\sim m_T v_X^2$ (for $M_X>> m_T$), where $m_T$ is the target mass, $M_X$ is the dark matter mass and $v_X$ is the velocity of the incident dark matter. Since the target mass can be assumed to be the mass of a typical nucleus in the human body, such as oxygen, and the velocity of dark matter in our galaxy can be assumed to be on average $250$~km/s, we get a recoil energy of $E_R \sim 10$~keV. On the other hand, due to its weak interaction with ordinary matter, the scattering rate of a typical WIMP-like dark matter particle is of order $\sim \mathcal{O}(10)$ scattering events per year with the human body, which is equivalent to a radiation dose of $\mathcal{O}(10^{-11})$~mSv/yr \cite{Freese:2012rp}. In comparison~\cite{ParticleDataGroup:2012pjm}, cosmic ray muons, a few of them passing through our body every second, deposit each $\sim 10 - 100$~MeV, corresponding to a radiation dose of below $0.8$~mSv/yr. An instant dose of radiation of $4$-$5$ Sv is considered lethal. The allowed exposure for radiation workers is $50$ mSv per year (in the U.S.). 

Recently, the WIMP assumption has come under pressure since dark matter has not appeared in dark matter search experiments designed to look for WIMP dark matter, and the community is increasingly doubting the WIMP hypothesis. This has led physicists to speculate that dark matter may have much stronger interaction and be much heavier than previously thought. In principle, dark matter could be as heavy and strongly interacting as a bullet, and with typical dark matter velocities in our galaxy of $250$~km/s it would be a lethal encounter for humans. The absence of such unexplained death or serious injury was recently used by \cite{Sidhu:2019oii} to constrain a previously unexplored parameter space for dark matter. It was shown that the absence of such unexplained impacts with a well-monitored subset of the human population could be used to exclude a region bounded by a dark matter cross-section $\sigma_X\geq10^{-8}$-$10^{-7}$cm$^2$ and a dark matter particle mass $1$ mg $<M_X<50$ kg. 

If, however, dark matter is lighter and with weaker interaction strengths than considered by Sidhu, Scherrer and Starkman, but still very strongly interacting and very heavy compared to a WIMP, it would not lead to immediate death or serious injury. Instead, we want to investigate if it could amount to significant radiation exposure due to its ionizing properties. 

The interesting cross-sections would require dark matter to be in a composite state \cite{Digman:2019wdm}, which can have either an elastic or inelastic scattering cross-section with ordinary nuclei. We will consider both cases in turn, first the elastic scattering and then the more model-dependent inelastic case.

\section{Elastic scattering}

\begin{center}
\begin{figure}[t]
\includegraphics[width=\textwidth]{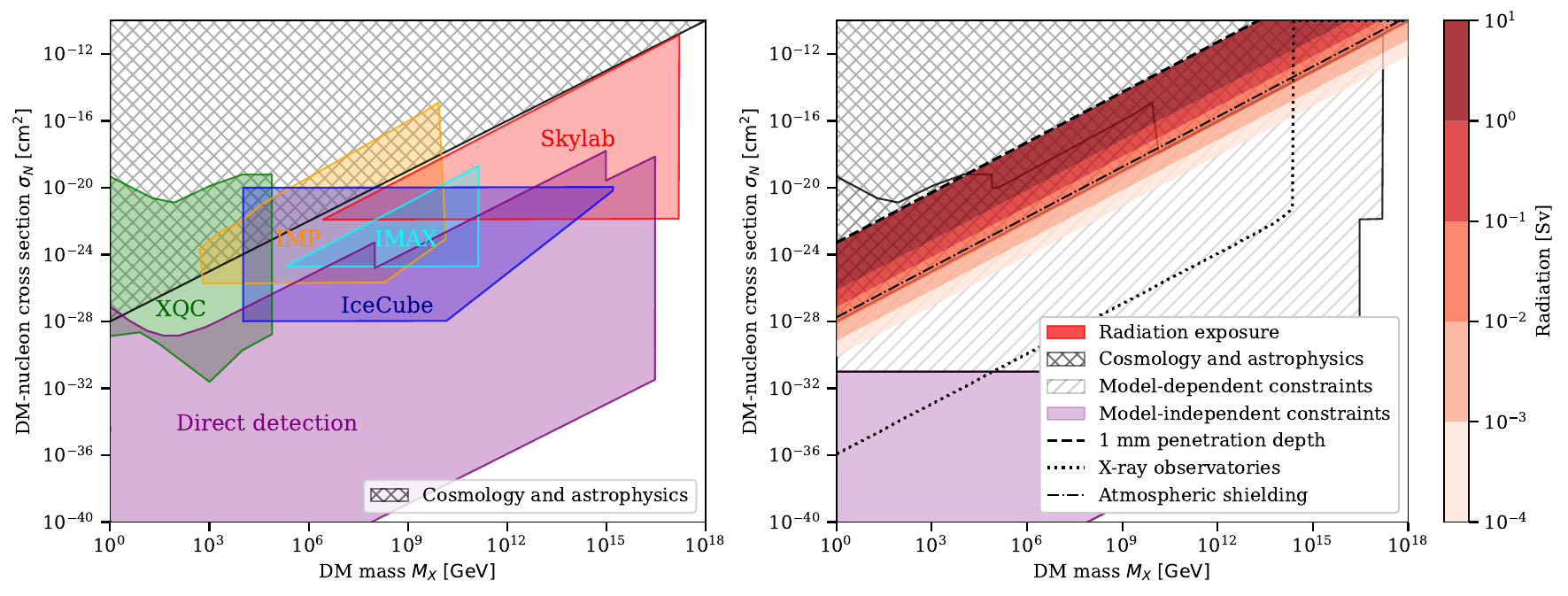}
\caption{Constraints on the spin-independent dark matter nucleon cross-section $\sigma_N$ as a function of dark matter mass~$M_X$. \underline{Left panel:} Current constraints on the parameter space (adopted from \cite{Digman:2019wdm,Kavanagh:2017cru,Wandelt:2000ad}). Except for the direct detection constraints below the threshold $\sigma_N \sim 10^{-31} \mathrm{cm}^2$, all constraints are model-dependent~\cite{Digman:2019wdm}, either because they assume the scaling relation $\sigma_X \propto A^4 \sigma_N$, which does not hold in general, or they require further assumptions (such as self-annihilation into neutrinos in the case of IceCube). For example, in~\cite{Bhoonah:2020fys}, the Skylab constraint is discussed when assuming a different scaling of A for contact interactions or making more conservative assumptions about perceived background events, which could be misidentified as strongly interacting dark matter events. \underline{Right panel:} We display the radiation exposure in Eq.~(\ref{exposure}) for $f = 1$ in terms of an annual whole body dose (gradually red-shaded band). Below the dash-dotted line atmospheric shielding can be neglected and a small exposure could be observed on Earth, above it a more significant radiation dose could affect humans on prolonged space travel. The dashed line represents an upper limit above which the radiation cannot penetrate human skin (or the shielding of the space craft). Our scenario is particularly relevant in cases where the cosmological bound does not apply, though an allowed region remains even when it does. }
\label{fig:elastic}
\end{figure}
\end{center}

Using the known dark matter density and dark matter velocity, we can estimate the dark matter flux here on Earth as a function of the dark matter particle mass to be\footnote{To be specific, we assume $v_X=250 \mathrm{km}/\mathrm{s}$ and $\rho_X = 3 \times 10^5 \mathrm{GeV}/\mathrm{m}^3$. } 
\beq
F_{X}\simeq 7.5 \times 10^{6}\textrm{cm}^{-2}\textrm{s}^{-1}f\left(M_X/\textrm{GeV}\right)^{-1}\, ,
\eeq
 where $f$ is the fraction of this type of dark matter relative to the total dark matter.  
 Similarly, we can estimate the mean free path of a dark matter particle in the human body as a function of the dark matter nucleus cross-section $\sigma_X$ to be 
\beq\label{eq:lambda}
\lambda_X = \mu/(\rho_h \textrm{N}_a \sigma_X) = 3\times 10^{-23}\textrm{cm}^3/\sigma_X\, , 
\eeq
where $\rho_h\approx 1$g/cm$^3$ is the approximate human mass density, $\mu$ is the molar mass of water, and N$_a$ is Avogadro's number\footnote{To capture the main effect for spin-independent scattering, we take the oxygen nuclei as the targets since they lead to a higher energy deposit than scattering off a single proton~(see also ~\cite{Freese:2012rp} for a more careful treatment of different nuclei in the human body).}. Now, since $E_{R} \approx 20~\mathrm{keV}$ of energy is deposited in the human body per collision, we have that the energy deposited in a human per length is 
\beq\label{dEdl}
\mathrm{d} E/\mathrm{d}x \approx 20 \textrm{keV}/\lambda_X\,.
\eeq
The depth of a human is approximately $\Delta X =10$ cm, so the energy deposited in a human in one year ($\Delta t = 1$ yr) is on average 
\beq
\Delta E = \Delta X (\mathrm{d}E/\mathrm{d} x) F_X A \Delta t\approx 7 \times 10^{37}\textrm{GeV}f(\sigma_X/\textrm{cm}^2)(M_x/\textrm{GeV})^{-1}\, ,
\eeq
where we took the area of a human to be $A=1.7\times 10^4$ cm$^2$ \cite{Sparreboom}. Here, we arrive at a subtle point. With $\sigma_X$, we denote the cross section for scattering dark matter off a nucleus in the body. To compare our results to existing bounds in the literature, we need to convert $\sigma_X$ to the cross section on nucleons $\sigma_N$. However, as pointed out in \cite{Digman:2019wdm}, the often used scaling relation $\sigma_X \propto A^4 \sigma_N$ (with $A$ the atomic mass number) is not applicable for cross sections $\sigma_X > 10^{-31} \mathrm{cm}^2$, rendering the discussion model-dependent. Specifically, this failure is traced back to the breakdown of the first Born approximation when the naive cross-section $\sigma_X \propto A^4 \sigma_N$ would exceed the geometric cross-section. To obtain the correct constraints, one would need to specify a concrete (composite) dark matter model and compute $\sigma_X$ explicitly. For the purposes of this work, we will parametrize this uncertainty through $\sigma _X = w_N \sigma_N$, where $w_N > 1$, and for our numerical estimates, we will simply take $w_N$ to be of order unity.\footnote{In fact, discarding the scaling of the cross-section with $A$ when plotting constraints at large cross-sections is also advocated for in \cite{Digman:2019wdm} as the most correct approach when dark matter is composite.} This is a conservative approach because it underestimates for a given value of $\sigma_N$ the expected radiation exposure (because the actual dark matter nucleus cross section is larger). We can now ask what is the dark matter cross-section that would expose a human to a whole-body dose of radiation $\Delta S$ in one year. We find\footnote{Note, the radiation weighting factor, $w_R$, which converts between Grays and Sievert, is always greater or equal unity, $w_R \geq 1$ \cite{Donahue:2004aq}.}
\beq \label{exposure}
\frac{\sigma_N}{\textrm{cm}^2} \approx  10^{-28} w_R^{-1} w_N^{-1} \left( \frac{0.1}{f}\right) \left(\frac{M_X}{\mathrm{GeV}}\right)\left(\frac{\Delta S}{\textrm{mSv}}\right)~.
\eeq
The corresponding radiation levels are indicated in right panel of Fig.~\ref{fig:elastic} as the red shaded contours. We note that this formula is only valid in the limit where the dark matter deposits a small fraction of its kinetic energy in the human body (otherwise, one would need an improved treatment). This translates into the upper bound
\begin{align}\label{sigmabound}
\frac{\sigma_N}{\mathrm{cm}^2} \lesssim   10^{-26} w_N^{-1} \left(\frac{M_X}{\mathrm{GeV}}\right)\,.
\end{align}
Above this bound, the radiation exposure will saturate to a value $ \Delta S \approx  6 \times w_R   f \, \mathrm{Sv}$ (corresponding to a complete energy conversion). This still allows for a sizable radiation dose. In particular, if $\Delta S > 1$ mSv, then the dark matter will be a larger source of radiation exposure than other sources of background radiation that humans are exposed to on Earth.  We also indicated by the dashed line in Fig.~\ref{fig:elastic} the upper limit of the cross-section above which the penetration depth in water is less than 1 mm, the thickness of the human skin, blocking any radiation and cutting of the red region. To estimate the upper bound on the radiation exposure here on Earth, we note that for cross-sections $\sigma_X \geq 10^{-28}$ cm $(M_X/\mathrm{GeV})$, the dark matter will lose a significant amount of its kinetic energy going through the atmosphere~\cite{Starkman:1990nj}. Dark matter above the dash-dotted line in Fig.~\ref{fig:elastic} will, therefore, not be an immediate threat to life on Earth and cannot be detected in experiments on Earth. Setting $f=1$ in Eq.~(\ref{exposure}), we see that we obtain as a conservative estimate a radiation dose that could be as large as $10$ mSv here on Earth, although the real upper bound could be higher (due to the uncertainties in $w_R$ and $w_N$). 
Finally, we only consider cross-sections $\sigma_N \lesssim 10^{-12} \,\mathrm{cm}^2$ as dark matter will lose its ionizing property for larger values.

These values for the dark matter cross-section and mass fall into a regime that has been tested extensively by different experiments. Indeed, by looking at the left panel in Fig.~\ref{fig:elastic} one might easily conclude that there is no remaining parameter regime where dark matter could lead to a sizeable radiation exposure.  Here, the XQC (green)~\cite{Erickcek:2007jv}, IMP (orange), IMAX (cyan) and Skylab (red) constraints rely on atmospheric and space-based detectors~\cite{Wandelt:2000ad}. The hashed contour summarizes cosmology constraints~\cite{Nadler:2019zrb,Buen-Abad:2021mvc}, and the purple contour arises from direct detection experiments~\cite{Kavanagh:2017cru}. However, as we will argue now, this conclusion might be premature.
 As pointed out in~\cite{Digman:2019wdm} (and mentioned above), most of the colored constraints in Fig.~\ref{fig:elastic} use the conversion rule $\sigma_X \propto A^4 \sigma_N$, which is unreliable for $M_X \gtrsim 1 \mathrm{GeV}$ and $\sigma_N \gtrsim 10^{-31} \mathrm{cm}^2$.  Overall, without the $A^4$ enhancement, constraints should be expected to become weaker. Moreover, the IceCube constraint is trivially model-dependent because it assumes that dark matter annihilates into neutrinos~\cite{Albuquerque:2010bt}. We, therefore, toned down these model-dependent constraints in the right panel of Fig.~\ref{fig:elastic}.\footnote{Underground detectors {(purple region in Fig.~\ref{fig:elastic} adopted from~\cite{Digman:2019wdm}, see also \cite{Kavanagh:2017cru}}) are not sensitive to these cross-sections, as they are shielded against cosmic radiation, which will also shield them against this type of dark matter.}\footnote{We have included the most generic constraints but there are additional model-dependent constraints that we have not considered, see for example~\cite{Ray:2023auh}.}
With the large elastic cross-sections we are interested in, our scenario would also affect the way dark matter clumps and seeds structure in the early universe, and therefore, it would alter the measured spectrum of the Cosmic Microwave Background (CMB) anisotropies and structure formation. Since for cosmological probes $A=1$, the above scaling complication would not arise. The strongest cosmological bound comes from the Milky Way Satellites (MWS) and Lyman-alpha data, whereas the CMB bound is weaker~\cite{Nadler:2019zrb,Buen-Abad:2021mvc, Hooper:2022byl}. These constraint do have some additional model dependence due to late-time baryonic effects, which are modeled by mapping constraints from Warm Dark Matter (WDM) simulations to the interacting case \cite{Nadler:2019zrb}. There could be unknown redshift dependence in dark matter properties or baryonic feedback processes \cite{Tulin:2017ara}, as well as self-interactions in the dark matter sector, which could lead to enhanced small-scale structure \cite{Fan:2013yva}, possibly compensating for the damping of the small-scale structure due to the dark matter nucleon interaction. Another possibility is that only a small fraction $f$ of dark matter has formed large composite states, like dark nuclei or atoms, relaxing the cosmological bounds. To be specific, the Lyman-alpha bound relaxes as $f^3$ and the weaker CMB bounds as $f$ \cite{Hooper:2022byl}. We assume, like in \cite{McKeen:2022poo}, that the MWS bound will also relax linearly with $f$. It is, therefore, plausible that there is a remaining parameter window, unconstrained by more conventional dark matter experiments, for which there could exist a significant radiation exposure not only in space but also on Earth. 

Of course, one might expect that this radiation would have been picked up by other experiments, for example, geared towards detecting cosmic radiation on earth, and therefore should be treated as likely excluded.\footnote{However, the energy threshold of cosmic ray detectors is typically higher than what is needed to detect elastic dark matter with similar mass and cross sections, due to the lower kinetic energy of dark matter particles.} A careful study of the exact limits provided by known types of cosmic ray detectors and dosimeters for measuring ionizing radiation could, in principle, provide interesting new limits on dark matter properties. Another example is provided by X-ray experiments.
Since the recoil energy of the nucleus is in the range of typical X-ray telescope sensitivities of $0.1-10$ keV, we have indicated a new tentative constraint with the dotted contour in Fig.~\ref{fig:elastic} where the energy deposit of a dark matter particle going through the detector of an  X-ray telescope is above its typical sensitivity threshold, like the Chandra X-ray telescope\footnote{See also \url{https://www.chandra.harvard.edu/about/specs.html}.} \cite{Marshall:2003ir}, of $10^{-15}$erg/s/cm${}^2$. For $\sigma_N\gtrsim 10^{-21}$cm${}^2$, the dark matter particle will scatter multiple times in a detector with a thickness of $\sim 1$ mm, the thickness of a CCD chip, in a time much shorter than the 16-microsecond time resolution of the High-Resolution Camera of the Chandra X-ray telescope. Therefore, for dark matter with  $\sigma_N\gtrsim 10^{-21}$cm${}^2$, the energy deposit of a single event is much above the $10$ keV upper threshold of a single X-ray photon, and the event would possibly be registered as a ``pile-up event", for which the sensitivity bound is more uncertain. Thus, the excluded region above the dotted line in  Fig.~\ref{fig:elastic} should be considered with additional caution for  $\sigma_N\gtrsim 10^{-21}$cm${}^2$. One may also ask if this type of dark matter could be detected by attempts to measure the radiation exposure of astronauts in space. However, the energy threshold of this type of detector is typically too high. For example, the detection threshold for the detector in~\cite{phantom} is approximately $10^5 \mathrm{keV}/\mathrm{cm}$. The general radiation monitoring device on the International Space Station (ISS), ISS-RAD \cite{2023LSSR...39...67Z}, as well as the LIDAL Time-of-Flight Radiation Detector on the ISS \cite{2023Senso..23.3559R}, specially designed for identifying the particle types of ionizing radiation, have both a threshold of order of $10^4$ keV/cm. Using \eqref{eq:lambda} and \eqref{dEdl}, this translates into the upper bound $\sigma_N < 10^{-20} \mathrm{cm}^2/ w_N$, below which the energy loss per length, $\mathrm{d}E/\mathrm{d}x$, is too small to exceed the dosimeter and radiation detector thresholds\footnote{The M-42 dosimeter of the RadMap experiment on the ISS, whose data from 2023 are now being analyzed, has an energy threshold of $60$ keV with a chip thickness of $300$ $\mu$m, which translates into a slightly better potential sensitivity of 2000 keV/cm \cite{Losekamm:2023prt}.}.  Nevertheless, even above that threshold, heavy and strongly interacting dark matter could contribute an important radiation component for long-duration space travel, such as missions to Mars (provided we can ignore the model-dependent cosmology constraints). 

This opens up the possibility that space-based radiation experiments might be used to detect or constrain dark matter in the future. As one concrete example, a Radiation Assessment Detector (RAD), with a similar threshold as ISS-RAD, was originally designed for the Mars Science Laboratory (MSL), a robotic space probe mission to Mars. On the way to Mars in 2012, it measured radiation exposure in a thinly shielded spacecraft in deep space corresponding to $0.4$ Sv per year \cite{2013Sci...340.1080Z,rad}. It would, therefore, be interesting to more carefully analyze the constraints these space-based radiation experiments would impose on the red contour in Fig.~\ref{fig:elastic}.

One can imagine many possible composite dark matter states with large enough masses and cross-sections, which could, in principle, contribute a noticeable radiation exposure to space travelers.  Some simple examples are dark nucleons and nuclei (for a review, see \cite{Kribs:2016cew}). Here, we can, for simplicity, just consider a bound state of bosons described by the Lagrangian
\beq
\mathcal{L} \supset g_{\chi}m_\chi\phi\chi^*\chi +g_N\phi \bar\psi_N \psi_N  \,,
\eeq
where $\chi$ is the dark matter scalar field and $\psi_N$ is the one related to the nucleon. The scalar field $\phi$ is a mediator with mass $\mu\geq 1$ TeV that couples to $\psi_N$  with strength $g_N$ through a Yukawa interaction. When integrated out the effective dark matter nucleon interaction is described by a contact operator:
\beq
\mathcal{O}_c \sim \frac{g_\chi g_N m_\chi \chi^*\chi \bar\psi_N \psi_N}{\mu^2}~.
\eeq
In this model, one can have bound states of $\chi$ particles of mass $M_X$ consisting of $N_X = M_X /m_\chi$ particles, each of mass $m_\chi$, similar to the dark blobs discussed in \cite{Grabowska:2018lnd}. {Following their work, it can be checked that it is consistent with having dark matter blobs with mass $10^5\, \mathrm{GeV}\lesssim M_X \lesssim 10^{18} \,\mathrm{GeV}$ for 1 TeV$\lesssim \mu \lesssim 10^{16}$ GeV and $m_\chi\sim 0.1 \times\Lambda_\chi$, where 1 keV $\lesssim\Lambda_\chi\lesssim 1$ MeV is the Bohr radius of the bound state's constituents. The cross-section of these dark blobs with nucleons will then be given by the geometric cross section $\sigma_N \sim \pi \Lambda_\chi^{-2}$ in the range $ 10^{-20}$cm$^2 \lesssim \sigma_N \lesssim 10^{-14}$ cm${}^2$.} From the above discussion, we see that such dark matter blobs would deposit almost all their energy and could potentially lead to up to $\Delta S \approx  6 \times w_R   f \, \mathrm{Sv}$ of whole-body radiation exposure per space traveler per year and thus a dose 1000 times the background radiation on Earth if the model-dependent MWS constraint does not apply. This is far above the radiation exposure, which is considered safe, and opens the hypothetical possibility that dark matter can be an important source of radiation exposure for space travelers and, to a much weaker extent, also to humans on Earth.

\section{Inelastic scattering}

If dark matter scatters inelastically, it can deposit an even larger amount of energy per collision while being significantly harder to detect. An example is electrically neutral Q-balls (as opposed to the charged Q-ball discussed in the previous section), which have also widely been considered as a dark matter candidate \cite{Kusenko:1997si,Kusenko:1997vp}. {However, to keep our discussion as general and model-independent as possible, we will simply think of a dark blob, similar to that discussed in the previous section, but with constituents with $B-L$ violating interactions, which can catalyze baryon destruction as discussed for single-particle states in \cite{Ema:2024wqr}.}

The constraint on the elastic scattering of heavy dark matter with the human body mostly applies also to inelastic scattering but with a few important exceptions. Consider the process where one has electrically neutral dark matter that interacts inelastically with a nucleon by destroying the nucleon and ripping it apart in a process like
\beq
{ X + p^+ \to \tilde X + e^+ +\pi \dots} \,,
\eeq
where the interaction of the dark matter particle $X$, with the proton $p$, creates an excited dark matter state $\tilde X$, a positron and some pions. We can then assume that the excited dark matter state continues through the medium, interacting with more protons. At the same time, the positron and pions will also move through the medium with a mean free path of typically several centimeters. Thus, typically, the {positron and pions} will interact again electromagnetically with the medium a few centimeters away from the original interaction point of dark matter. The track of the neutral inelastic heavy dark matter will, therefore, be delocalized to a region of the width of a few centimeters and not carve out a sharp track in the material as looked for by the Mica experiment~\cite{Snowden-Ifft:1995zgn}. The Mica constraint does, therefore, not apply to this form of dark matter. In addition, the effect on humans will be more subtle, and the constraint of \cite{Sidhu:2019oii} from shooting holes in human bodies must be reevaluated in this case. In the case considered above, where the inelastic interaction of dark matter with nucleons produces a positron and pions, all the energy of the positron and pions is absorbed in the body. 

On the other hand, due to the inelastic nature of the scattering, the dark matter blob loses a negligible part of its energy in collisions with matter particles on its way through the atmosphere and the Earth.
This also means that underground detectors are not shielded, and the biggest one, IceCube, gives an absolute flux constraint of (using that not more than a few events could have gone unnoticed per year)
\beq
F_X \leq 10^{-17}\textrm{cm}^{-2} \textrm{s}^{-1}\, .
\eeq
 IceCube's area is $1~ \textrm{km}^2$. With the area of an average human being $1.7$ m$^2$, IceCube's area corresponds to the area of roughly 600,000 people. Since IceCube has not detected dark matter, after accounting for background noise and inefficiency in the detector, only a few such blobs can pass through IceCube per year. Or maximally, a few in 600,000 people can be hit by this type of dark matter per year. Over a person's life span of order 100 years, this implies that roughly 1 person in 1000 can be hit by a baryon destroying dark matter blob. 

On the other hand the flux of dark matter is {$F_{X}\simeq 7.5\times 10^{6}$cm$^{-2}$s$^{-1}f\left(M_X/\textrm{GeV}\right)^{-1}$}. The mass of a dark matter particle, which hits 1/1000 humans, is then $10^5 \, m_\mathrm{pl} /f$, where $m_\mathrm{pl}$ is the reduced Planck mass.
The energy deposit per length of a blob traversing a human body per nucleon collision is $\sim 1$ GeV, but due to the large geometric cross-section, the blob will interact with the entire nucleus. We will therefore assume that the energy deposit per nucleus collision is $\sim 10$ GeV, where the mean free path is as in the elastic case $\lambda_X \sim 3\times 10^{-23} \textrm{cm}^3/\sigma_X$. Now, since, in this case, $10$ GeV of energy is deposited in the human body per collision, we have that the energy deposited in a human per length is
\beq
\mathrm{d}E/\mathrm{d}x = 10  \, \textrm{GeV}/\lambda_X\,,
\eeq
and since the depth of a human is approximately $\Delta X =10$ cm, the energy deposited in a human hit by such a {blob} is   
\beq
\Delta E = \Delta X (\mathrm{d}E/\mathrm{d}x) \approx 0.3\times 10^{25}\, \textrm{GeV}(\sigma_X/\textrm{cm}^2)\, .
\eeq 
Thus, the cross-section that gives $\Delta S =1$ Sv of instant whole-body radiation when hit by a blob is
\beq
\frac{\sigma_X}{\textrm{cm}^2} = 2\times 10^{-13}\left(\frac{\Delta S}{\textrm{Sv}}\right)~.
\eeq
Thus, for baryon destroying dark matter blobs with a mass of $10^5 \, m_\mathrm{pl}$ and  $\sigma_X \simeq 10^{-13}$ cm$^2$, we conclude that 1/1000 humans could once in their lifetime experience an instant radiation dose of 1 Sv from a dark matter encounter. 

\section{Conclusion}

Given the unknown properties of dark matter, we have investigated the upper limit on how much radiation humans here on Earth or in space could be exposed to, in principle,  due to hypothetical forms of dark matter, which are not excluded by known direct or indirect dark matter searches.

Our study shows that for dark matter, which scatters elastically with the ordinary nucleons, the radiation dose could be at least as large as $10$ mSv per year, which is more than a factor 30 higher than that of known cosmic radiation. The atmosphere is an important shield against heavy, strongly, elastically interacting dark matter, but in space, humans are more exposed. In space, there are very few constraints on the potentially harmful effects dark matter could have. The MWS provides the strongest constraint, but it is model-dependent. If the MWS constraint is ignored, humans on prolonged space travels could hypothetically be exposed to up to $0.6$ Sv per year, which is 1000 times higher than on Earth.  It would, however, not be challenging to build dosimeters with a lower threshold, which could place stronger bounds on this form of dark matter. A version of the XQC experiment but with a much longer flight time {or a careful re-analysis of pile-up events in X-ray telescopes} could exclude this possibility. Alternatively, a careful re-analysis of previous space-based radiation experiments might equally allow the exclusion of part of this parameter regime.

If dark matter scatters inelastically with ordinary nucleons, one can not exclude the possibility that up to one per mille of the human population on Earth is exposed to an instant radiation dose of up to 1 Sv once in their lifetime due to dark matter interactions. This, however, requires that a small component of dark matter is in a heavy composite state with significant interactions with ordinary matter. While this possibility cannot at present be excluded, it would be interesting to investigate this in more depth to see if stronger bounds can be placed.

\subsection*{Acknowledgements}
We thank Torsten Bringmann, Mathias Garny, and Alexander Kusenko for their comments on our first draft. F.N. is supported by VR Starting Grant 2022-03160 of the Swedish Research Council. M.S.S. is supported by Independent Research Fund Denmark grant 0135-00378B. 

\bibliographystyle{JHEP}
\bibliography{ref}

\end{document}